%% file: ms.tex
\documentclass[12pt, a4paper]{article}
\input{packages}

\title{\bf On the relevance of the Godot Engine in the indie game development industry}
\author{Julian Holfeld, M. Sc.\\
University of Kassel \\
\href{mailto:julian.holfeld@morphclue.com}{julian.holfeld@morphclue.com}}
\date{}

\twocolumn
\begin{document}
\maketitle
\input{sections/abstract}
\input{sections/introduction}
\input{sections/godot}
\input{sections/methodology}
\input{sections/results}
\input{sections/discussion}

\bibliographystyle{unsrt}

\end{document}

%% file: packages.tex
\usepackage[english]{babel}
\usepackage[utf8]{inputenc}
\usepackage{csquotes}
\usepackage{titlesec}
\usepackage{graphicx}
\usepackage{hyperref}
\usepackage{fancyhdr}
\usepackage{tabularx}
\usepackage[left=2.5cm,right=2.5cm,top=2.5cm,bottom=2cm]{geometry}
\usepackage[dvipsnames]{xcolor}
\usepackage{xurl}
\usepackage{mathabx}

%% file: sections/abstract.tex
\section*{Abstract}
This paper examines the relevance of the Godot Engine in the indie game industry.
The Godot Engine is a relatively new game engine from 2014 and competes with leading market players.
To get to the bottom of its relevance, two major online sales platforms and the game engines that are commonly used, Steam and itch.io, are examined.
Mainly, these findings are compared with reference data from 2018.
It turns out that the Godot engine has gained massive relevance in 2020 and now seems to be one of the leading players in the indie game industry.
The exact causes are difficult to determine.
However, this paper provides many clues for further research in this area.

%% file: sections/introduction.tex
\section{Introduction}
In the realm of video game development, the methodology for creating games has evolved significantly over time.
As early as the 1980s, several companies began using their own in-house engines.
For instance, in the case of Super Mario Bros (1985), Nintendo used a fast scrolling game engine that was originally developed for Excitebike (1984) \cite{history-digital-games}.
The concept of licensing game engines to other companies as a standalone product has been around since the mid-90's.
At that time, games like Doom (1993) were developed using the id Tech game engine \cite{id-tech-doom}.
This engine was specifically designed for 3D shooters, although this iteration was not yet a fully-fledged 3D game engine.
Over the years, an increasing number of game engines were developed to speed up the development of video games.\\
 
It should be mentioned that the definition of a game engine differs in literature.
In this paper, game engines are seen as frameworks that accelerate the development process and provide at least support for 2D or 3D rendering.
It is crucial to recognize that not only does the definition differ significantly, but so does the use case of a game engine.
A prime example is the id Tech engine, which was developed explicitly for a particular genre of video games.
There are also game engines like Unity, which could be called a general purpose game engine.
Therefore, a comparison between game engines is not easy.
Since they always serve a specific purpose, there is no game engine that stands above all others.
This paper takes a closer look at the purpose of video game development in the indie game industry, with a particular focus on the Godot Engine.
The decision to investigate this specific game engine is driven by the insights derived from a few datasets, which will be expounded upon later in this paper.

%% file: sections/godot.tex
\section{Godot Engine}
The Godot Engine was released as open source under the MIT license on GitHub on February 10 in 2014\footnote{The first commit containing the source code can be found under the SHA:\\ 0b806ee0fc9097fa7bda7ac0109191c9c5e0a1ac} \cite{godot-repository}.
The first stable release was published in December 2014 \cite{godot-release}.
This makes it relatively new in comparison to popular game engines like Unity and Unreal Engine, which were released in 2005 and 1998, respectively \cite{unity-release, unreal-release}.
Considering the different release dates, it should be clear that due to the young age of the Godot Engine the ecosystem that formed around this technology is still small in comparison to older engines.\\

The Godot Engine supports cross-platform exports and facilitates the development of 2D and 3D games \cite{godot-engine}.
Video games can be exported for PC, mobile devices or web platforms.
These games are written with a programming language specially developed for the Godot Engine called GDScript.
GDScript is a dynamically typed programming language which is syntactically similar to Python.
However, the Godot Engine also officially supports C\texttt{\#} and includes GDNative, a technology for creating bindings to other languages \cite{godot-gdnative}.
These officially include C and C\texttt{++}.
Unofficially support for languages such as D, Kotlin, Nim, Python and Rust have been developed by the community.\\

It's noteworthy that the Godot Engine appears to attract not only smaller companies but also larger players like Microsoft, Epic Games and Meta's Reality Labs.
Microsoft donated \$24,000 to the Godot team in 2017 for implementing C\texttt{\#} as a programming language \cite{godot-csharp}.
In 2020, the Godot Engine received the Epic MegaGrant of \$250,000, which is distributed by Epic Games to projects that expand the open source 3D graphics ecosystem \cite{godot-megagrant}.
This is noteworthy as Epic Games develops the Unreal Engine, suggesting a potential interest in supporting other game engines for mutual benefit.
The adoption of an MIT license, as in the case of Godot Engine, may facilitate the incorporation of new technologies without negative implications for intellectual property.
In December 2020 and 2021 Godot Engine received donations from Meta's Reality Labs \cite{godot-facebook-reality, godot-meta-reality}.
The funds will be used specifically for development in the field of extended reality (XR).
There is no public information available regarding the exact amount of the donation.
This was confirmed in a Reddit comment by project manager Rémi Verschelde (akien-mga) \cite{reddit-companies-akien}.
Rémi also mentioned that the team actively approaches companies rather than the other way around.

%% file: sections/methodology.tex
\section{Methodology}
In the last section it became evident that the Godot Engine has managed to capture the attention of larger companies, suggesting a degree of relevance in the video game industry.
However, this does not directly address the central question of how relevant the Godot Engine is for indie game developers.
To effectively answer this question, we must first establish a clear definition of indie game development.
Just as the definition of a game engine is elusive, so is the definition of indie game development.
The definition used here is from O'Donnell, who states that indie games focus on a few goals and can be developed on a smaller scale in terms of manpower \cite{indie-definition}.
An indie game development team operates without financial backing from a larger company, distinguishing it from AAA titles developed with substantial budgets and large teams. \\

To investigate the relevance of the Godot Engine for indie game developers, a comprehensive literature review was conducted.
It is important to note that this research was not limited to scholarly articles but also included gray literature.
This is because the Godot Engine is a relatively new game engine, and there is a lack of scholarly articles on the topic.
The gray literature used in this paper was carefully selected to ensure reliable and trustworthy information. \\ 

For this paper, the digital sales platforms Steam and itch.io were selected for investigation.
Firstly, it aligns with a study conducted by Toftedahl and Engström that provides comparative data from 2018 \cite{game-engine-taxonomy}.
Secondly, reference is made to a survey named ``State of the Game Industry'' from 2023 encompassing 2,300 video game developers \cite{gdc-2023}.
Notably, the focus in this paper lies on PC and not on console or mobile games.
This is because the mentioned survey from 2023 showed that 65\% of the respondents are actively developing for PC and 57\% plan to do so in the next project as well.
In 2019, a survey involving nearly 4,000 video game developers highlighted that, aside from publisher-owned and individual websites, most games are primarily sold on Steam and itch.io \cite{gdc-2019}.\\

Identifying the game engine used in games available on Steam presents a significant challenge due to a lack of detailed information.
In the paper by Toftedahl and Engström mentioned above, a script was used that matches games with articles on Wikipedia and subsequently assigns game engines.
However, this procedure presents challenges and gaps in the data sets.
This is because not all games provide information about the underlying game engine. \\

Alternatively, another method involves recognizing the game engines used by analyzing filenames through pattern matching.
The website SteamDB employs this method to deduce the game engines utilized in development \cite{steamdb-tech}.
While this approach provides valuable insights, it is not without its limitations, including the potential for gaps or false positives due to ambiguous filenames.
In this paper, the datasets from SteamDB are used because matching filenames to game engines yield more accurate results than matching the data to Wikipedia articles.
In addition, the different versions of the id Tech engine were combined as one. \\

Analyzing the data from itch.io proves to be more straightforward because the website provides a list of game engines used in development with the number of games for each engine.
This structured presentation of data facilitates ease of access and analysis.
This approach aligns with the methodology employed by Toftedahl and Engström in their research.
However, it is important to acknowledge that this method captures a static snapshot of the data at a given point in time.
To address this limitation a script was developed that extracts data from itch.io's RSS feed in XML format, specifically recording the publication date of each game \cite{github-trend-itch}.
This additional temporal dimension allows for the creation of a time series, enabling a more in-depth analysis of used game engines over time.

%% file: sections/results.tex
\section{Results}
When examining the presence of various game engines on Steam for 2018, 2022, and 2023, it is important to see the significance of these results in the context of indie game development.
The following \autoref{table:steam} presents the percentages of game engines on Steam for these years.\\

\begin{table}[ht!]
    \resizebox{\columnwidth}{!}{
        \centering
        \begin{tabular}{|l c c c|}
            \hline
            Game Engine & 2018   & 2022    & 2023                                                \\
            \hline\hline
            Unity       & 25.6\% & 61.22\% & $\textcolor{red}{\blacktriangledown 60.04\%}$       \\
            Unreal      & 13.2\% & 15.91\% & $\textcolor{ForestGreen}{\blacktriangleup 16.42\%}$ \\
            Source      & 4.0\%  & 0.27\%  & $\textcolor{red}{\blacktriangledown 0.23\%}$        \\
            CryEngine   & 3.5\%  & 0.23\%  & $\textcolor{red}{\blacktriangledown 0.19\%}$        \\
            GameBryo    & 3.2\%  & -       & -                                                   \\
            IW          & 2.9\%  & -       & -                                                   \\
            Anvil       & 2.5\%  & 0.05\%  & $\textcolor{gray}{\bullet 0.05\%}$                  \\
            id Tech     & 1.7\%  & 0.18\%  & $\textcolor{red}{\blacktriangledown 0.15\%}$        \\
            Essence     & 1.1\%  & -       & -                                                   \\
            Clausewitz  & 1.0\%  & 0.03\%  & $\textcolor{red}{\blacktriangledown 0.02\% }$       \\
            Other       & 48.4\% & 22.17\% & $\textcolor{ForestGreen}{\blacktriangleup 22.94\%}$ \\
            \hline
        \end{tabular}
    }
    \caption{Percentage of total games identified on Steam (data collected 2022-09-30 and 2023-10-21)}
    \label{table:steam}
\end{table}

Game engines that have experienced an increase in their percentage of total games developed in comparison to the previous year(s) are highlighted in green, indicating a positive trend, while those that have seen a decrease are highlighted in red, suggesting a decline in their usage over the same period.
The data from 2018 was collected using the Wikipedia method, while the data from 2022 and 2023 was collected using the pattern matching method.
Color-coding in \autoref{table:steam} is missing for the column for 2022.
This is due to the differing data collection methods used in 2018.
It is difficult to say whether the newer method can attribute more game engines or simply more games were created using a particular game engine.
To illustrate, consider Unity, which has 35.62\% more share of all identified games from 2018 to 2022.
Additionally, it is worth noting that among the game engines analyzed, only the Unreal Engine has consistently exhibited growth over the various years.
This trend holds true when considering both the complete dataset, which includes 2018 data, and the dataset from 2022 and 2023 alone.
The Godot Engine is not listed in \autoref{table:steam} because the paper from Toftedahl and Engström did not include it in their analysis.
Despite this omission, the data remains valuable in shedding light on changing market dynamics. \\

Between 2022 and 2023, it is noteworthy that other game engines have also gained increased popularity, indicating shifts in the competitive landscape.
At the time of the collected data in 2022, 1.15\% of games were created with the Godot Engine.
However, in 2023, this percentage increased to 1.44\%.
Compared to other game engines on SteamDB, the Godot Engine ranks 6th out of 65 recognized game engines.
In addition to that it is important to understand that the percentages from \autoref{table:steam} represent lower bounds.
Particularly in the case of the Godot engine, there could be a lot more games.
This is due to the fact that the Godot Engine allows for the export of executables without a recognizable filename signature, making them unidentifiable by the SteamDB algorithm.
Consequently, only a subset of games with specific filename patterns can be matched. \\

\begin{figure}[ht!]
    \begin{center}
        \includegraphics[width=1\columnwidth]{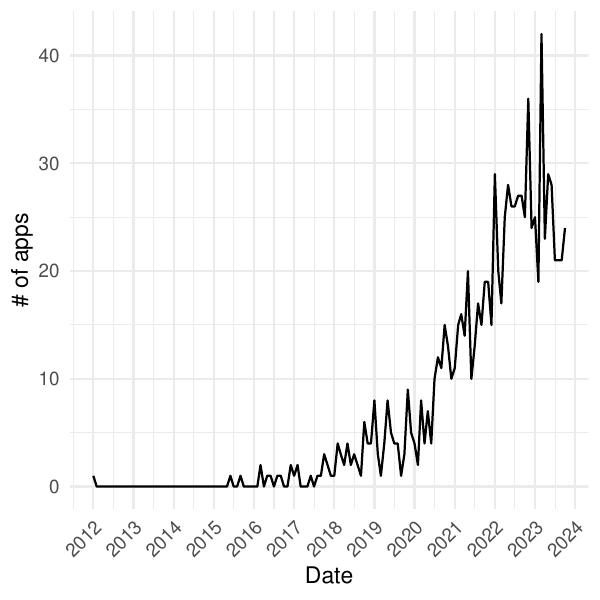}
        \caption{\label{fig:godot-graph} Number of released Apps on Steam per month}
    \end{center}
\end{figure}

Examining the monthly app releases depicted in \autoref{fig:godot-graph}, a clear trend emerges: in 2020, there was a significant increase in the number of released games.
This increase in game releases prompts further investigation to understand the underlying reasons.
Prior to conducting a more in-depth analysis of the underlying reasons, we first draw a parallel by comparing the data from itch.io.
In contrast to Steam, the digital sales platform itch.io predominantly showcases indie games, aligning with the specific focus of this paper.
Unlike Steam, itch.io possesses an official statistics website that provides information on game engines \cite{itchio-engines}.
However, it's essential to acknowledge that this data might be incomplete due to its reliance on self-reporting.
Therefore, there may be gaps or inaccuracies in the information, a limitation inherent to self-reported data. \\

\begin{table}[ht!]
    \resizebox{\columnwidth}{!}{
        \centering
        \begin{tabular}{|l c c c|}
            \hline
            Game Engine & 2018   & 2022                                                & 2023                                                \\
            \hline\hline
            Unity       & 47.3\% & $\textcolor{ForestGreen}{\blacktriangleup 49.75\%}$ & $\textcolor{red}{\blacktriangledown 46.33\%}$       \\
            Construct   & 12.3\% & $\textcolor{ForestGreen}{\blacktriangleup 13.12\%}$ & $\textcolor{ForestGreen}{\blacktriangleup 13.82\%}$ \\
            GameMaker   & 11.0\% & $\textcolor{red}{\blacktriangledown 7.32\%}$        & $\textcolor{red}{\blacktriangledown 6.95\%}$        \\
            Twine       & 6.2\%  & $\textcolor{red}{\blacktriangledown 5.35\%}$        & $\textcolor{ForestGreen}{\blacktriangleup 6.03\%}$  \\
            RPG Maker   & 3.9\%  & $\textcolor{red}{\blacktriangledown 2.74\%}$        & $\textcolor{ForestGreen}{\blacktriangleup 2.76\%}$  \\
            Bitsy       & 3.3\%  & $\textcolor{red}{\blacktriangledown 3.11\%}$        & $\textcolor{ForestGreen}{\blacktriangleup 3.18\%}$  \\
            PICO-8      & 2.9\%  & $\textcolor{red}{\blacktriangledown 2.68\%}$        & $\textcolor{red}{\blacktriangledown 2.60\%}$        \\
            Unreal      & 2.8\%  & $\textcolor{ForestGreen}{\blacktriangleup 2.92\%}$  & $\textcolor{ForestGreen}{\blacktriangleup 3.01\%}$  \\
            Godot       & 2.5\%  & $\textcolor{ForestGreen}{\blacktriangleup 5.55\%}$  & $\textcolor{ForestGreen}{\blacktriangleup 7.51\%}$  \\
            Ren'Py      & 2.0\%  & $\textcolor{red}{\blacktriangledown 1.93\%}$        & $\textcolor{ForestGreen}{\blacktriangleup 2.07\%}$  \\
            Other       & 5.9\%  & $\textcolor{red}{\blacktriangledown 5.55\%}$        & $\textcolor{ForestGreen}{\blacktriangleup 5.75\%}$  \\
            \hline
        \end{tabular}
    }
    \caption{Percentage of total games identified on itch.io (data collected 2022-09-30 and 2023-10-21)}
    \label{table:itch}
\end{table}

Similar to Steam, the data from itch.io can be compared to the 2018 reference paper using the most current data available from 2022 and 2023.
The consistent data collection method allows for direct comparisons, as illustrated in \autoref{table:itch}.
Upon comparing the data presented in \autoref{table:steam} and \autoref{table:itch}, Unity stands out as a prevalent game engine.
In both 2022 and 2023, Unity consistently constitutes over half of all games on Steam and nearly half on itch.io.
This dominance is particularly striking when considering the second-ranking game engines, which represent approximately 16\% of games on Steam and roughly 14\% on itch.io in 2023.
A similar trend can be observed for 2022. \\

It can also be seen that only three game engines have consistently gained popularity on itch.io.
These three game engines are Construct, Unreal and the Godot Engine.
Construct is a game engine primarily aimed at non-programmers.
With the help of visual programming, it is intended to make it easier for novice programmers to get started.
The notable increase in its percentage of games may be attributed to the appeal of game development to a broader audience with limited coding expertise.
However, this is only an assumption that needs to be investigated further. \\

Looking at \autoref{table:itch}, it becomes evident that the Godot Engine has experienced remarkable growth in popularity on itch.io.
To delve deeper into this trend, the previously mentioned script was used to compile games from itch.io along with their release dates.
It should be mentioned that not all games could be fetched with this script.
Despite the missing data sets of up to about 10\% of the games, it is still possible to create a trend plot for the game engines listed in \autoref{table:itch}.

\begin{figure}[ht!]
    \begin{center}
        \includegraphics[width=1.1\columnwidth]{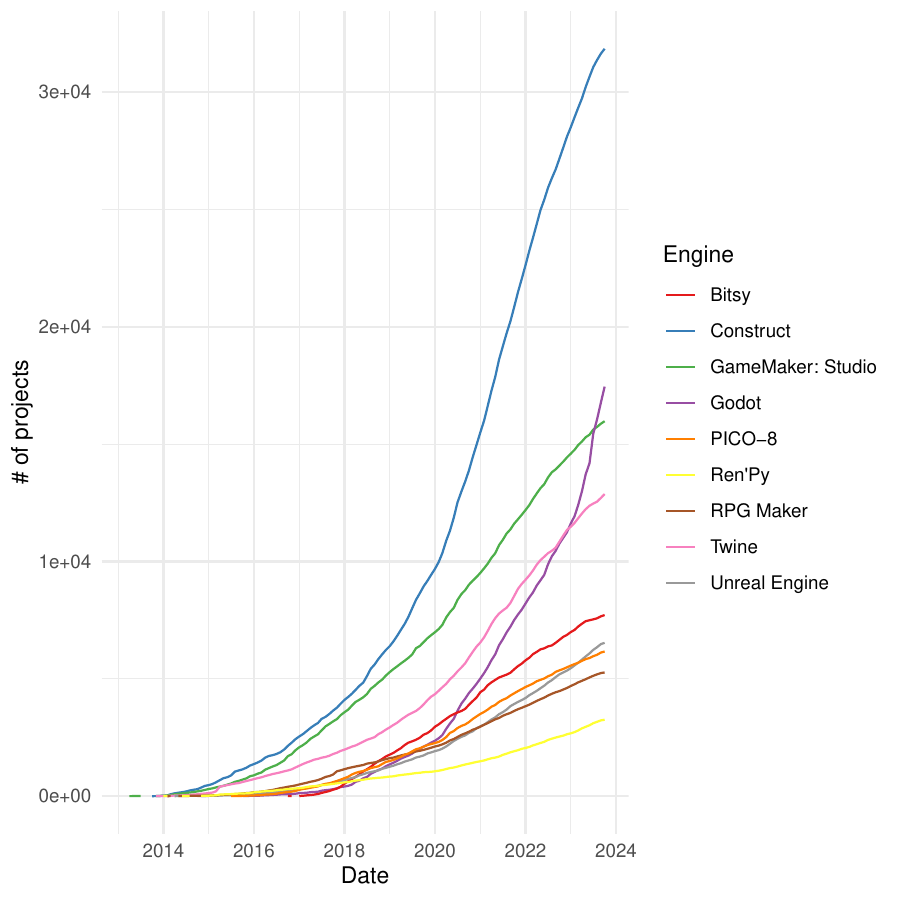}
        \caption{\label{fig:trend-graph} Cumulative sum of the number of projects on itch.io}
    \end{center}
\end{figure}

This graph is shown in \autoref{fig:trend-graph}.
Notably, Unity, due to its large number of projects, has been excluded from the graph to ensure better visibility and clarity for the other game engines.
It is evident that the Godot Engine has gained a significant surge in popularity in 2020, surpassing both Bitsy and Twine.
This analysis has its limitations, as it relies on the quantity of games published.
Quality and quantity of games by individual publishers were not considered.
Determining the exact percentage of indie game studios is challenging as not every publisher was individually analyzed.
\newpage

\begin{table}[ht!]
    \resizebox{\columnwidth}{!}{
        \centering
        \begin{tabular}{|l c c c c|}
            \hline
            Game Engine & 2020   & 2021                                               & 2022                                               & 2023                                             \\
            \hline\hline
            Unity       & 62.2\% & $\textcolor{red}{\blacktriangledown 61.6\%}$       & $\textcolor{red}{\blacktriangledown 61.1\%}$       & $\textcolor{red}{\blacktriangledown 59\%}$       \\
            Godot       & 12.2\% & $\textcolor{ForestGreen}{\blacktriangleup 13.1\%}$ & $\textcolor{ForestGreen}{\blacktriangleup 15.6\%}$ & $\textcolor{ForestGreen}{\blacktriangleup 19\%}$ \\
            GameMaker   & 10.9\% & $\textcolor{red}{\blacktriangledown 8.9\% }$       & $\textcolor{red}{\blacktriangledown 6.1\%}$        & $\textcolor{red}{\blacktriangledown 5\%}$        \\
            Unreal      & -      & 4.2\%                                              & $\textcolor{ForestGreen}{\blacktriangleup 4.8\%}$  & -                                                \\
            Construct   & 1.7\%  & $\textcolor{ForestGreen}{\blacktriangleup 2.4\%}$  & $\textcolor{red}{\blacktriangledown 1.7\%}$        & -                                                \\
            Stencyl     & 0.1\%  & 0.1\%                                              & -                                                  & -                                                \\
            Other       & 12.9\% & $\textcolor{red}{\blacktriangledown 6.5\%}$        & $\textcolor{ForestGreen}{\blacktriangleup 7.0\%}$  & $\textcolor{ForestGreen}{\blacktriangleup 17\%}$ \\
            No engine   & -      & 3.1\%                                              & $\textcolor{ForestGreen}{\blacktriangleup 3.7\%}$  & -                                                \\
            \hline
        \end{tabular}
    }
    \caption{Used game engines by GMTK Game Jam participants over the years \cite{gmtk-twitter}}
    \label{table:gmtk}
\end{table}

An interesting distinction on itch.io, which is not as prominent on Steam, is the prevalence of games created during game jams, which are defined as follows:
\blockquote{A game jam is an accelerated opportunistic game creation event where a game is created in a relatively short timeframe exploring given design constraint(s) and end results are shared publically \cite{game-jam-definition}.}
With around 18,000 - 23,000 participants and around 6,000 - 7,000 submissions, GMTK Game Jam is the largest past game jam event on itch.io \cite{gmtk-game-jam-2021, gmtk-game-jam-2022, gmtk-game-jam-2023}.
\autoref{table:gmtk} shows the game engines used by participants in the events from 2020 to 2023.
This table is clearly showing that the Godot Engine is the only game engine from the table to outperform the usage of the previous year three years in a row. \\

To explore more about the Godot Engine, the Godot Community Poll from 2022 and 2023 was examined \cite{godot-poll-results-22,godot-poll-results-23}.
In 2022, there were 5,315 responses to the survey, while in 2023 the number increased to 7,671.
These surveys contain answers to a variety of questions.
Regarding previous experiences with other game engines, respondents in the 2022 survey mentioned Unity, other third-party engines, and GameMaker.
In the 2023 survey, the trends remained consistent.
The surveys show that most people heard about the Godot Engine for the first time between 2018 and 2020.
However, most did not start development with the engine until the following years 2019 to 2022.
The surveys also clearly show that the game engine is mainly used for 2D game development.\\

The survey from 2022 sheds light on further facts that are relevant to indie game development.
A substantial 84\% of participants stated that they use the Godot Engine as a hobby, while 9\% identified themselves as full-time indie game developers.
According to the definition outlined in this paper, hobby developers fall under the category of indie game developers, as they share common characteristics such as focusing on specific goals, lacking financial support from larger companies, and working in small teams.
This is further supported by the finding that 97.9\% work on their project alone or in a team of up to five people.
Furthermore, the survey reveals that 83.7\% of respondents do not earn any income from their games, emphasizing their indie status.
Additionally, the two most prevalent digital sales platforms among respondents are itch.io (7.1\%) and Steam (6\%).

%% file: sections/discussion.tex
\section{Discussion}
When examining various data sets, it becomes clear that the popularity of the Godot Engine has increased.
Starting with the digital sales platform Steam, it is noteworthy that the Godot Engine ranks 6th out of 65, which indicates its importance even at lower bounds, which arise due to data collection.
For a more meaningful discussion, Unity and Unreal Engine will also be included in this section.
According to \autoref{table:steam} Unity lost a few percent from 2022 to 2023 while Unreal Engine gained a few percent.
A possible reason for this could be that AAA titles are also published on Steam and Unreal Engine is known to be used for developing such games \cite{ue-reinforcement-learning}.\\

On the other hand, this argument explains why growth on a platform like itch.io for Unreal Engine is rather slow compared to the Godot Engine.
While the Unreal Engine prominently appears as the second most used game engine on Steam, its presence on itch.io has seen only marginal growth of roughly 0.2\% over the past four to five years.
Furthermore, \autoref{table:itch} shows that during the same period of time, the Godot Engine has gained a share of roughly 5\%.
This is particularly interesting as it suggests that the Unreal Engine has limited relevance among indie developers, while the Godot Engine has become increasingly relevant in the indie game sector over time.
This observation is further supported by the fact that the Godot Engine ranks third on itch.io.\\

Although Unity is the biggest player in the indie games sector, \autoref{table:itch} clearly shows that other game engines are gaining popularity for indie game development.
Another game engine under competitive pressure is GameMaker, which has lost roughly 4\% of its share over the years.
This observation can also be reproduced in the context of game jams, where Unity is gradually being displaced by other competitors, while GameMaker has been rapidly losing ground in recent years.
In contrast, the Godot Engine has seen a significant increase in popularity in game jams, where it has become the second most used game engine in the GMTK Game Jam.
It is important to consider that the overall popularity of a game engine could lead to its increased utilization in game jams.
Otherwise, it should also be noted that some games published in the GMTK Game Jam end up on itch.io and in this way also increase the number of games developed with the Godot Engine.
Further insight would require a precise analysis of this interaction. \\

Unity and GameMaker will persist as references in the Godot Engine Community Polls, as they are mentioned in relation to participants who have experience with these game engines.
Giving the declining shares of Unity and GameMaker, it is possible to assume that certain developers may have switched entirely to the Godot Engine.
There could be many reasons for this.
On one hand, the license of the Godot Engine offers great freedom and on the other hand, certain features could be favored.
However, in order to draw a direct comparison, a deeper analysis would also have to be carried out.
It should also be noted that not every game engine can be compared with every other, as mentioned at the beginning.
Each game engine has its own purpose and therefore a comparison is not easy.
If a comparison is nevertheless desired, one possible useful comparison would be between Unity and the Godot Engine, as both can be classified as general-purpose game engines that cover a wide range of game genres.\\

The results of this study make it clear that the Godot Engine definitely has relevance in the indie game industry.
In all three data sets analyzed (Steam, itch.io and GMTK Game Jams), the Godot Engine ranks better than most game engines.
In addition, it can also be said that various larger companies that have shown an interest in the Godot Engine through financial support indicating its relevance in the industry. 
The only question that remains is why this is the case.
Both \autoref{fig:godot-graph} and \autoref{fig:trend-graph} clearly show 2020 as a successful year for the Godot Engine.
If you combine this finding with the answers from the Godot Community Poll, you can see that people had already heard of the game engine before 2020, but only developed with it later.
It is also quite possible that events before 2020 were decisive for the rapidly growing popularity in the indie game industry.
However, this data reveals little about the reasons for this rise. Even after extensive research, it is difficult to pinpoint the exact reason for the growth.
This is probably due to the fact that many factors are responsible for this growth. \\

Another insight is that most games developed with the Godot Engine are mainly 2D games.
This is also in line with the statements from the Godot community survey.
On one hand, the Godot Engine might offer more robust features for 2D game development compared to its 3D capabilities.
On the other hand, the amount of work involved in developing 3D games is usually higher than that of 2D games, which can be a challenge for indie developers, causing them to favor 2D development. \\

The methods of analysis used here have raised many new questions.
These should continue to be investigated.
The market relevance in general or specifically for indie game studios could be further analyzed.
Possible suggestions include interviews with indie game studios that use the Godot Engine.
Additionally, in the case of Unity and GameMaker, it would be interesting to do further research to see if the Godot Engine is slowly pushing them out of the market.\\

One suggestion is to interview indie game studios that use the Godot Engine to gain further insights.
Furthermore, direct comparisons, where appropriate, would be useful to compare technical aspects between the Godot Engine and other game engines.
Methods such as a feature matrix, benchmarks or alternative methodologies would be useful for this. 
Although the reasons for Godot Engines growth in popularity are unclear, it can be concluded that the Godot engine is highly relevant in the indie game industry and will probably continue to grow over the next few years.